\documentclass[]{aastex631}

\usepackage{tcolorbox}
\tcbuselibrary{listings}

\begin{document}

\title{AstroLLaMA-Chat: \protect\includegraphics[height=3em]{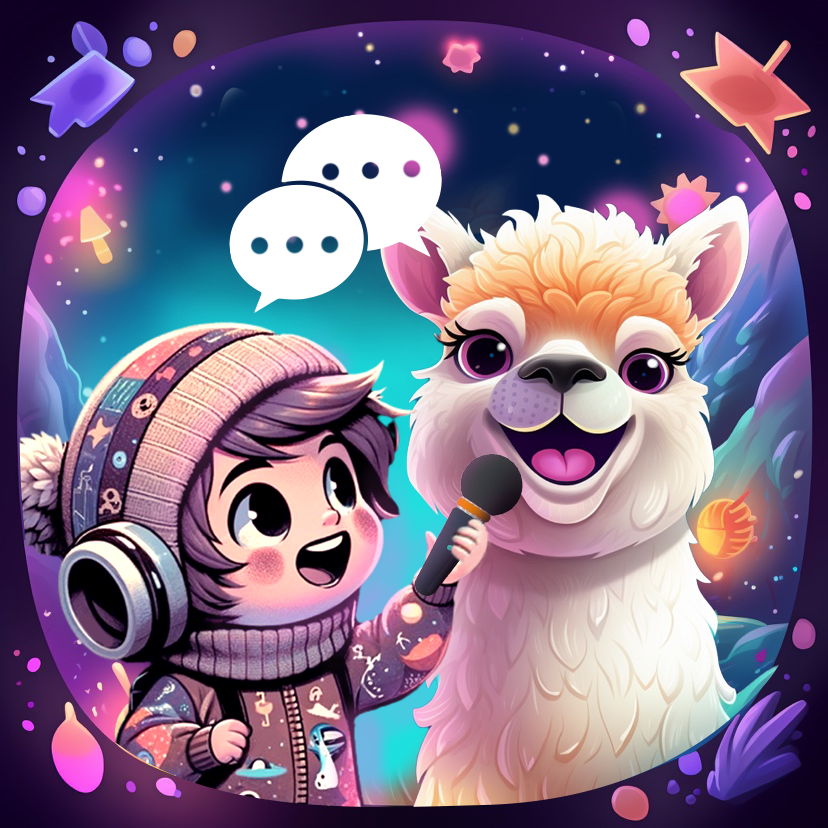} Scaling AstroLLaMA with Conversational and Diverse Datasets}

\author{Ernest Perkowski$^*$}
\affiliation{European Space Agency (ESA), European Space Astronomy Centre (ESAC), Camino Bajo del Castillo s/n 28692 Villanueva de la Cañada, Madrid, Spain}

\author{Rui Pan$^*$}
\affiliation{
Department of Computer Science and Engineering, Hong Kong University of Science and Technology}

\author{Tuan Dung Nguyen}
\affiliation{Department of Computer and Information Science, University of Pennsylvania, Philadelphia, PA 19104, USA}

\author{Yuan-Sen Ting}
\affiliation{Research School of Astronomy \& Astrophysics,
Australian National University, Cotter Rd., Weston, ACT 2611, Australia}
\affiliation{School of Computing, Australian National University, Acton, ACT 2601, Australia}
\affiliation{Department of Astronomy, The Ohio State University, Columbus, OH 43210, USA}
\affiliation{Center for Cosmology and AstroParticle Physics (CCAPP),
The Ohio State University, Columbus, OH 43210, USA}

\author{Sandor Kruk}
\affiliation{European Space Agency (ESA), European Space Astronomy Centre (ESAC), Camino Bajo del Castillo s/n 28692 Villanueva de la Cañada, Madrid, Spain}

\author{Tong Zhang}
\affiliation{Department of Computer Science, University of Illinois Urbana-Champaign}   

\author{Charlie O'Neill}
\affiliation{Mathematical Science Institute, Australian National University, Acton, ACT 2601, Australia}

\author{Maja Jablonska}
\affiliation{Research School of Astronomy \& Astrophysics,
Australian National University, Cotter Rd., Weston, ACT 2611, Australia}

\author{Zechang Sun}
\affiliation{Department of Astronomy, MongManWai Building, Tsinghua University, Beijing 100084, China}

\author{Michael J. Smith}
\affiliation{Aspia Space, Tremough Innovation Centre, Penryn TR10 9TA, United Kingdom}

\author{Huiling Liu}
\affiliation{Department of Modern Physics, University of Science and Technology of China, Hefei, Anhui 230026, China}

\author{Kevin Schawinski}
\affiliation{Modulos, Technoparkstrasse 1, 8005 Zurich, Switzerland}

\author{Kartheik Iyer}
\affiliation{Columbia Astrophysics Laboratory, Columbia University, New York, NY 10027, USA}

\author{Ioana Ciucă}
\affiliation{Research School of Astronomy \& Astrophysics,
Australian National University, Cotter Rd., Weston, ACT 2611, Australia}
\affiliation{School of Computing, Australian National University, Acton, ACT 2601, Australia}

\author{UniverseTBD}

\def\thefootnote{*}\footnotetext{These authors contributed equally to this work.}

\begin{abstract}

\noindent We explore the potential of enhancing LLM performance in astronomy-focused question-answering through targeted, continual pre-training. By employing a compact 7B-parameter LLaMA-2 model and focusing exclusively on a curated set of astronomy corpora---comprising abstracts, introductions, and conclusions---we achieve notable improvements in specialized topic comprehension. While general LLMs like GPT-4 excel in broader question-answering scenarios due to superior reasoning capabilities, our findings suggest that continual pre-training with limited resources can still enhance model performance on specialized topics. Additionally, we present an extension of AstroLLaMA: the fine-tuning of the 7B LLaMA model on a domain-specific conversational dataset, culminating in the release of the chat-enabled AstroLLaMA for community use. Comprehensive quantitative benchmarking is currently in progress and will be detailed in an upcoming full paper. The model, AstroLLaMA-Chat, is now available at \url{https://huggingface.co/universeTBD}, providing the first open-source conversational AI tool tailored for the astronomy community.
\end{abstract}

\section{Motivation}
\label{sec:intro}

Large Language Models (LLMs) have demonstrated exceptional capabilities across a wide range of tasks, covering both general and specialized domains, as evidenced by models like GPT and LLaMA \citep{radford2019gpt2,brown2020gpt3,touvron2023llama,touvron2023llama2}. Despite their impressive achievements, these models face notable challenges in highly specialized fields such as astronomy, particularly in keeping abreast of the latest field developments. This limitation arises from two primary factors: firstly, LLMs’ propensity to align with general concepts restricts their capacity for providing detailed, nuanced responses in question-answering scenarios; secondly, infrequent updates to their training datasets result in a delay in assimilating recent astronomical advancements.

\section{AstroLLaMA-Chat}

Building upon our earlier initiative, AstroLLaMA \citep{nguyen2023astrollama}, the pioneering LLM tailored for astronomy and trained on over 300,000 arXiv paper abstracts using the LLaMA-2-7b model \citep{touvron2023llama2}, we identified that while AstroLLaMA excelled in abstract completion, its ability in question-answering tasks is still wanting. To enhance this, we introduce AstroLLaMA-Chat, an advanced version of AstroLLaMA. This new iteration broadens the training scope to include introductions and conclusions of papers, alongside abstracts, as these sections are often rich in pivotal information for question-answering tasks. We initiated by downloading all papers up to July 2023, including all the files that come with a submission to arXiv. The data has been further refined for optimal operability, retaining only files with “.tex” suffixes. Through a multi-stage process, and utilising a comprehensive regex matching process, the extraction of the targeted sections was performed. Given the diverse LaTeX formatting standards, approximately 90\% of the samples remained post-processing. Subsequently, we removed specific formatting patterns, comments, and superfluous symbols like newlines to ensure the readability of the training data.

Further, we have fine-tuned AstroLLaMA-Chat on a domain-specific dialogue dataset. To generate Question-Answer pairs, we engaged GPT-4 \citep{openai2023gpt4} to formulate pertinent questions from paragraphs within 300,000 arXiv papers, with GPT-4 also tasked with answering these questions by retrieving context-relevant information. This approach facilitated the extraction and conversational structuring of the dataset's knowledge, laying the groundwork for training a conversational bot. We created 10,356 samples from the abstracts of the aforementioned arXiv papers and integrated additional open-source datasets. The training involved a diverse mix of datasets, including the LIMA dataset \citep{zhou2023lima}, 10,000 samples from Open Orca \citep{OpenOrca,mukherjee2023orca,longpre2023flan}, and 10,000 samples from UltraChat \citep{ding2023enhancing}. 

\section{Training}

We executed fine-tuning on the LLaMA-2 models using the LMFlow LLM-training framework \citep{diao2023lmflow}, incorporating advanced techniques like Flash Attention \citep{dao2022flashattention,dao2023flashattention2}, ZeRO Optimization \citep{rajbhandari2020zero} and long-context techniques~\citep{chen2023extending}. This approach led to a significant efficiency gain: LMFlow reduced the time cost for training AstroLLaMA \citep{nguyen2023astrollama} by approximately fivefold. When training AstroLLaMA-Chat, despite the expanded dataset, LMFlow required only about twelve A100 GPU days. Here the domain-specific training takes $\ge 90\%$ of the time and the downstream chat training only takes $\sim 10$ GPU hours. Our hyperparameters for domain-specific training included a peak learning rate of \(\eta = 10^{-5}\), a batch size of 32, a maximum token length of 2048, a warmup ratio of 0.03, two gradient accumulation steps, and the use of the \texttt{fp16} format, while the downstream chat training shares a similar set of hyperparameters except for learning rate $\eta = 2\times 10^{-5}$ and batch size $64$. We opted for a cosine learning rate \citep{loshchilov2016cosine}, as empirical evidence suggested it enhanced model performance for our specific tasks. While we are releasing the 7b conversational model, our team has also trained a more substantial 70b version. The training for this larger model is considerably more resource-intensive, requiring approximately 160 A100 GPU days per epoch. We plan to release this 70b version in an upcoming full paper.

\begin{figure*}
  \centering
   \includegraphics[scale=0.72]{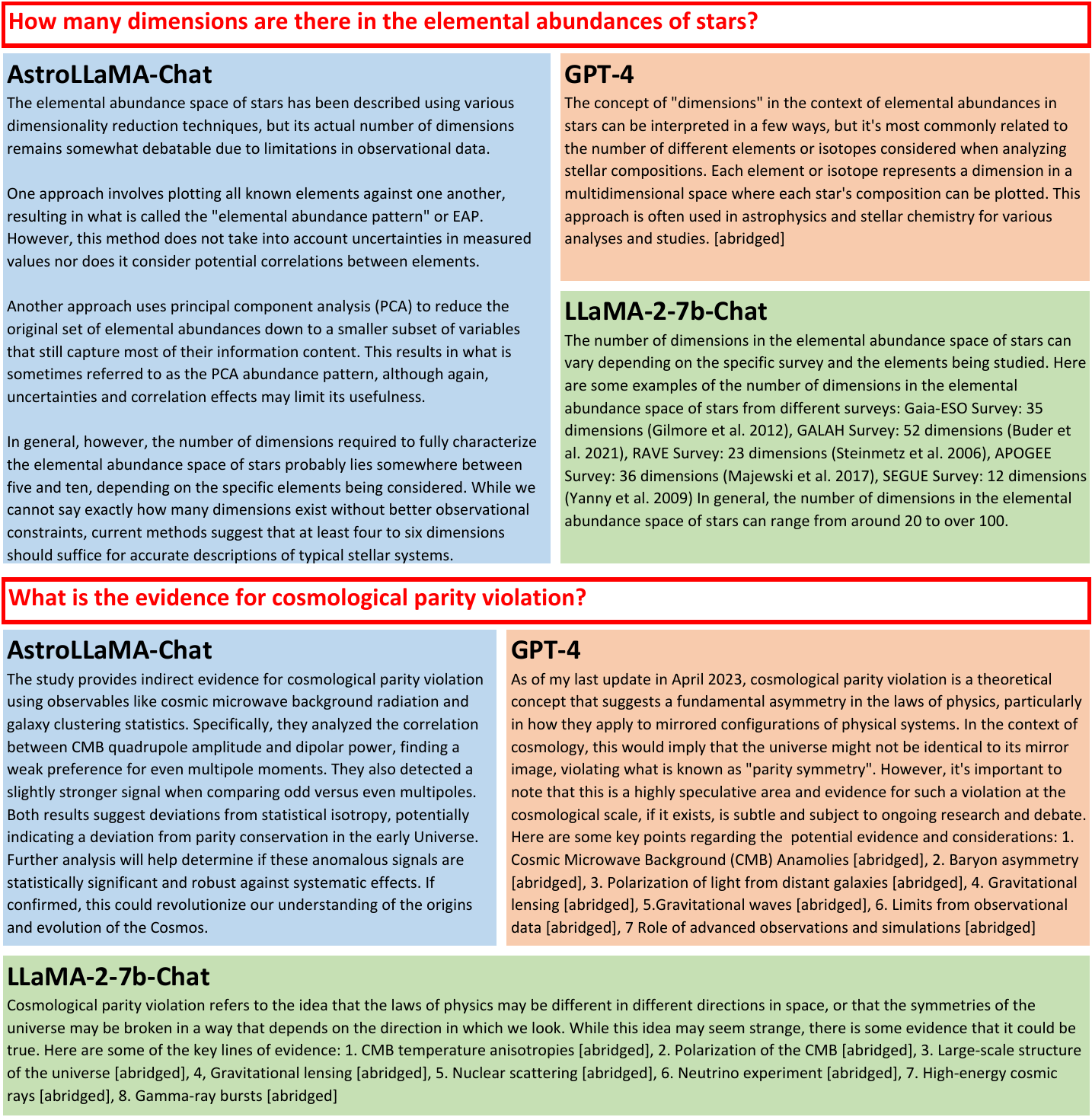}
   
  \caption{Demonstration of AstroLLaMA-Chat's Capabilities. While general large language models like GPT-4 continue to exhibit robust reasoning and Q\&A abilities, even in specialized domains such as astronomy, our study highlights the benefits of continual pre-training on a dedicated astronomy corpus from arXiv, enriched with the latest data. This approach gives AstroLLaMA-Chat an edge in two specific areas. The top example illustrates its performance in a highly specialized topic within astronomy. AstroLLaMA-Chat demonstrates a better understanding of the complexities involved in studying the dimensionality of elemental abundance in stars, reflecting the true chemical yield channels. It also outlines prevalent methods in this specialized area. In contrast, GPT-4 and the LLaMA-2-7b model, from which AstroLLaMA is derived, often provide responses that lack depth in understanding this field. The bottom panel illustrates AstroLLaMA-Chat's adeptness in addressing contemporary and dynamic research areas, notably the burgeoning field of parity violation studies in cosmology. While it captures some of the latest directions in the field (though with occasional detail inaccuracies), both GPT-4 and LLaMA-2 tend to diverge into broader implications and detection methods, failing to encapsulate the current focus of the field.}
  \label{fig:demo}
\end{figure*}

\section{Discussion}

A question naturally arises in the era of versatile and powerful large language models: Is there merit to developing specialized chatbots? Our findings indicate that general-purpose models such as GPT-4 and, to some extent, LLaMA-2, demonstrate robust reasoning and a good general understanding of astronomy. This suggests that with strategic prompting and engineering, existing large language models can serve as effective tools in this domain.

However, the primary objective of our research is to demonstrate that continual pre-training, even with a relatively modest model such as the 7b AstroLLaMA, can yield competitive and, in certain specific cases, superior performance. Our experiments reveal that while AstroLLaMA-Chat may not consistently outperform GPT-4 and LLaMA-2 in general astronomy-related Q\&A, it performs better in highly specialized topics. These include intricate areas like the dimensionality of elemental abundance space, differential line-by-line spectroscopic analysis, and recent studies in astronomy, such as the Large Magellanic Cloud (LMC) wake in the Milky Way's stellar halo or the cosmological parity violation. In these niche areas, AstroLLaMA tends to provide more accurate answers that GPT-4 and LLaMA-2, albeit still with limitations in alignment and a propensity for more hallucination. We aim to address the limitation of multi-turn conversations by enhancing our model in the future. This involves incorporating additional training data and implementing alignment techniques.

In addition to these specialized topics, AstroLLaMA-Chat, akin to what we have shown in AstroLLaMA-1\citep{nguyen2023astrollama}, shows a marginal edge in completing abstracts in astronomy articles, a feature now extended to introductions and conclusions. By contrast, LLaMA-2 occasionally deviates from its assigned tasks and is prone to errors. GPT-4, while sometimes providing informative responses, often generates overly lengthy answers that may not align well with the conventional format of a journal article in astronomy. For instance, given a prompt ``\texttt{Complete the following abstract: "Recent advances in X-ray binaries"}'' or ``\texttt{Recent advances in X-ray binaries}'', LLaMA-2-7b-Chat normally provides abstracts with different prefixes, while LLaMA-2-7b just generates empty or nonsense completions like ``\texttt{[Jonathan](https://github.com/jonathan-m)}'' from time to time. In comparison, AstroLLaMA-Chat outputs a rather reasonable completion about $3\times$ shorter than GPT-4 with the special prompt of  
``\texttt{\#\#\#ABSTRACT: Recent advances in X-ray binaries}''. This tendency of concise completion in AstroLLaMA-Chat can be attributed to its training procedure focus on reducing perplexity in causal completion. However, it is important to note that the improvements in the 7b model are somewhat modest. A more detailed quantitative analysis, including comparisons with our trained 70b models, will be presented in the full paper.

We hope this research note will inspire more astronomers to explore the fine-tuning of smaller models, achievable with modest computational resources (around 10 GPU days). Additionally, we are releasing these models on the Hugging Face demo playground. In a later version, this platform will allow  users to rate the responses with a thumbs up or down (\url{https://huggingface.co/spaces/universeTBD/astrollama-7b-chat-alpha}), offering valuable feedback from expert users. Such input is crucial as it will help advance this field of study, which, while still in its nascent stages, is already showing promising results.

\section*{Acknowledgements}
The authors thank Microsoft Research for their support through the Microsoft Accelerating Foundation Models Academic Research Program. We are also thankful for the support from OpenAI through the OpenAI Researcher Access Program.

\bibliography{ms}{}
\bibliographystyle{aasjournal}

\end{document}